\documentclass[11pt]{article}

\usepackage[usenames,dvipsnames,table]{xcolor}
\usepackage[utf8]{inputenc}

\pdfoutput=1 
\usepackage{jheppub} 

\usepackage{graphicx}
\usepackage{amsmath, amssymb}

\usepackage{mathrsfs}

\usepackage{framed}

\definecolor{shadecolor}{rgb}{0.90,0.90,0.90}

\newcommand{\qPoc}[2]{\left(#1;#2\right)_\infty}
\newcommand{\GF}[1]{\Gamma_e\left(#1\right)}
\newcommand{\thf}[1]{\theta_p\left(#1\right)}
\newcommand{\pq}[2]{(pq)^{\frac{#1}{#2}}}
\newcommand{\pqm}[2]{(pq)^{-\frac{#1}{#2}}}

\newcommand{\be}{\begin{eqnarray}}
\newcommand{\ee}{\end{eqnarray}}
\newcommand{\bea}{\begin{eqnarray}}
\newcommand{\eea}{\end{eqnarray}}

\newcommand{\nn}{\nonumber}
\newcommand{\bn}{\begin{enumerate}}
\newcommand{\en}{\end{enumerate}}

\def\Tr{\mathop{\mathrm{Tr}}\nolimits}

\def\SU{\mathrm{SU}}
\def\SO{\mathrm{SO}}

\def\U{\mathrm{U}}

\def\th{\theta}

\def\G{\Gamma}

\def\p{\partial}

\def\U{\text{U}}
\def\SU{\text{SU}}

\def\SO{\text{SO}}

\def\U{\text{U}}
\def\SU{\text{SU}}

\def\SO{\text{SO}}

\def\A{\text{A}}
\def\BC{\text{BC}}

\def\th{\theta_p}
\def\G{\Gamma_e}

\title{Ground state wavefunctions of elliptic relativistic integrable Hamiltonians}

\preprint{}

\author[a]{Belal Nazzal,}

\author[b]{Anton Nedelin,}

\author[a]{Shlomo S. Razamat }

\affiliation[a]{Department of Physics, Technion, Haifa, 32000, Israel}

\affiliation[b]{Section de Math\'ematiques, Universit\'e de Gen\`eve, 1211 Gen\`eve 4, Switzerland}

\emailAdd{sban@campus.technion.ac.il, anton.nedelin@gmail.com, razamat@physics.technion.ac.il}

\abstract{We derive ground state eigenfunctions and eigenvalues of various relativistic elliptic integrable models.
The models we discuss appear in computations of superconformal indices of four-dimensional theories obtained 
by compactifying six-dimensional models on Riemann surfaces. These include, among others, the Ruijsenaars-Schneider model and the 
van Diejen model.  The derivation of the eigenfunctions builds on physical inputs, such as conjectured 
Lagrangian across dimensions IR dualities and assumptions about the behavior of the indices in the limit of compactifications
on surfaces with large genus/number of punctures/flux.
}

\begin{document} 

\maketitle
\flushbottom

\section{Introduction} 

Elliptic integrable quantum mechanical systems are ubiquitous in the study of supersymmetric quantum field theories \cite{ Gorsky:1995zq,Donagi:1995cf,Nekrasov:2009rc,Gaiotto:2012xa,Nekrasov:2015wsu,Costello:2017dso,Costello:2018gyb,Costello:2018txb}. 
Typically these systems appear while accounting for various  protected sectors of such theories. It would be thus extremely interesting to understand 
the eigenvalues and eigenfunctions of such integrable systems. See for example
\cite{MR2472038,MR2506158,Razamat:2013qfa,MR4040584,MR4405564,vandiejen2021elliptic,Mironov:2021sfo,Awata:2020yxf,Awata:2020xfq,Awata:2019isq,noumi2012direct}.
In particular, as we will review below, studying the superconformal index of classes of
supersymmetric theories in four dimensions, the knowledge of the eigenfunctions leads to determination of the index even if a usual Lagrangian definition 
of a theory is not known.  The way the relations between the eigenfunctions and indices of four dimensional theories proceeds is through realizations of the $4d$ QFTs as compactifications of six dimensional SCFTs and derivation of the precise map between  geometric compactification data of a six dimensional theories and the four dimensional theories. See \cite{Razamat:2022gpm} for a review.

Many examples of such a map are known when we also have a different, independent, definition of the four dimensional theory: namely we have an across dimensions infrared duality.
In this note we will show how even a limited knowledge of a class of across dimensions dualities can lead to determination of eigenfunctions of certain elliptic integrable systems. On one hand we have a definition of the index of a class of theories using the Lagrangian in terms of sequences of elliptic hypergeometric integrals. On the other hand, under certain assumptions, the same index admits expansion in terms  of eigenfunctions of an elliptic integrable system. Making these assumptions and elementary arguments from statistical physics we can determine the eigenfunctions. The method  is particularly simple to apply to extract the eigenfunction of the ground state of the system: although the energies depend on complex parameters there is a natural ordering of the spectrum and a natural notion of the ground state. 

We will first outline the general method to determine the eigenfunctions in Section \ref{sec:meth}. We stress that the method has a physical input from across dimensional dualities and also relies on certain mathematical assumptions. 
 Then in Section \ref{sec:examples} we discuss several examples of applications of the method. First, we will consider the $\A_1$ Ruijsenaars-Schneider model. The eigenfunctions of this model are well known in certain limits of the parameters leading to {\it e.g} Macdonald polynomials. Here we will see how these can be easily computed perturbatively in parameters without taking any limits. The consistency of the result, at least in the perturbative expansion, gives evidence for the validity of our suggested eigenfunctions and the underlying conjectures.
 We will also discuss the $\BC_1$ van Diejen model and two somewhat more esoteric but simple integrable models which arose in physical contexts: we refer to these systems as the $\A_2$ and the $\A_3$ models. The paper is supplemented by a Mathematica notebook\footnote{\href{https://github.com/anedelin/GroundStates}{https://github.com/anedelin/GroundStates}} detailing all the reported computations.

\

\section{Ground state eigenfunction from large compactifications}\label{sec:meth}

We commence with a general discussion.
Let us consider the supersymmetric index \cite{Kinney:2005ej,Romelsberger:2005eg,Dolan:2008qi} of a compactification of some $6d$ $(1,0)$ SCFT $\color{blue}{T_{6d}}$ on geometry $\color{red}{\cal C}$ (defined by genus of a Riemann surface, the number and types of punctures, and flux for the $6d$ symmetry). For an ${\cal N}=1$ SCFT the superconformal index is defined as,
\be
{\cal I} = \Tr_{{\mathbb S}^3} (-1)^F \, q^{j_2-j_1+\frac{R}2} \, p^{j_2+j_1+\frac{R}2}\, \prod_{\ell=1}^{\text{rank}\, G_F} u_\ell^{{\cal Q}_\ell}\,.
\ee The trace is computed over the Hilbert space in radial quantization: that is quantization on ${\mathbb S}^3$ times the radial direction. Here $(j_1,\,j_2)$ are the two Cartan generators of the $su(2)\times su(2)$ isometry of ${\mathbb S}^3$; ${\cal Q}_\ell$ are Cartan generators of the global symmetry $G_F$; and $R$ is the charge under the superconformal R-symmetry.
The index is then a function of the compactification geometry as well as various parameters, 
\be 
{\cal I}[\textcolor{blue}{T_{6d}},\textcolor{red}{\cal C}](\{{\bf x}_j\},{\bf u}_{6d},q,p)\,.
\ee The  parameters are fugacities for various combinations of symmetries of the theory in $4d$. Parameters $q$ and $p$ are related to superconformal symmetry (as detailed above) and are there for any ${\cal N}=1$ SCFT in $4d$. The rest of the parameters correspond to various global symmetries. They depend on a theory at hand and are typically taken to be phases.
For theories arising in compactifications we can split the global symmetry $G_F$ into two kinds: the one coming from the symmetry, $G_{6d}$, of the $6d$ theory and the one associated to the punctures. Given the $6d$ SCFT the possibilities for the latter symmetry are classified as follows.  One first compactifies the $6d$ SCFT on a circle (possibly with holonomies for the $6d$ symmetry) and obtains an effective $5d$ description. In some cases the $5d$ description is in terms of a gauge theory and one could study boundary conditions  in the $5d$ spacetime. Such boundaries correspond to punctures and the various symmetries one can obtain are sub-groups of the $5d$ gauge group $G_{5d}$. Punctures with symmetry $G_{5d}$ are called maximal and these will play a special role for us. For a given $6d$ SCFT there might exist different circle compactifications leading to different $G_{5d}$ and thus to different kinds of maximal punctures.
In addition to being maximal or non-maximal, punctures can be distinguished by other properties.
We will collectively refer to maximal punctures with different defining properties as being of different types. From now on we will only turn parameters in the definition of the index corresponding to maximal punctures.

Given indices of two compactifications, each with at least two maximal punctures of the same type, one can compute the index of the compactification on a surface which is obtained by gluing the two surfaces along the two maximal punctures,
\be 
&&{\cal I}[\textcolor{blue}{T_{6d}},\textcolor{red}{{\cal C}_1\oplus {\cal C}_2}](\,\{{\bf x}_j\}_{S_1\cup S_2},{\bf u}_{6d},q,p)=\\
&&\,\oint d{\bf x} \;\Delta({\bf x}, {\bf u}_{6d};q,p)\;{\cal I}[\textcolor{blue}{T_{6d}},\textcolor{red}{{\cal C}_1}](\,\{{\bf x}_j\}_{S_1}\cup\{{\bf x}\},{\bf u}_{6d},q,p)\;
{\cal I}[\textcolor{blue}{T_{6d}},\textcolor{red}{{\cal C}_2}](\,\{{\bf x}_j\}_{S_2}\cup\{{\bf x}^{-1}\},{\bf u}_{6d},q,p)\,.\nonumber
\ee The parameters ${\bf x}$ correspond to the Cartan generators of $G_{5d}$ associated to the glued maximal punctures. The function
$\Delta({\bf x}, {\bf u}_{6d};q,p)$ is defined by properties of the punctures and is built from indices of various vector and chiral superfields one needs
to introduce when gluing the two punctures. The integration for each parameter is over the unit circle when we assume to take $|q|,\, |p|<1$.

Finally, given a $6d$ SCFT and a $5d$ circle reduction one obtains an elliptic relativistic integrable  model definined by a set (elements of which are parametrized by label $\alpha$) of commuting Hamiltonians \cite{Gaiotto:2012xa},
$${\cal H}_\alpha\left[\textcolor{blue}{T_{6d},G_{5d}}\right]({\bf x},{\bf u}_{6d};q,p)$$ such that indices corresponding to different compactifications are Kernel functions of these,
\be \label{Eq:kernel}
&&{\cal H}_\alpha\left[\textcolor{blue}{T_{6d},G_{5d}}\right]({\bf x}_1,{\bf u}_{6d};q,p)\cdot {\cal I}[\textcolor{blue}{T_{6d}},\textcolor{red}{\cal C}](\{{\bf x}_1,{\bf x}_2,\cdots\},{\bf u}_{6d},q,p)=\\
&&\;\;\;\;\;\; \qquad\qquad {\cal H}_\alpha\left[\textcolor{blue}{T_{6d},G_{5d}}\right]({\bf x}_2,{\bf u}_{6d};q,p)\cdot {\cal I}[\textcolor{blue}{T_{6d}},\textcolor{red}{\cal C}](\{{\bf x}_1,{\bf x}_2,\cdots\},{\bf u}_{6d},q,p)\,.\nonumber
\ee Here ${\bf x}_{1,2}$ correspond to maximal punctures of the same type and have $G_{5d}$ associated to them. Moreover, the Hamiltonians of the integrable system are self-adjoint under the scalar product defined using the measure $\Delta({\bf x}, {\bf u}_{6d};q,p)$.
The Hamiltonians ${\cal H}_\alpha$, when acting on the parameters corresponding to $G_{5d}$, introduce surface defects into the index computation.
The surface defects are labeled by $\alpha$ and the range of values for it depends on the theory at hand. The self-adjointness property follows from conjectured S-dualities that the underlying theories satisfy: {\it e.g.} the surface defect can be introduced by acting on any of the maximal punctures) \cite{Gaiotto:2012xa}. Because of the duality property the Hamiltonians for any choice of $\alpha$ commute. In practice, choosing a particular $6d$ theory to perform the computation for, one obtains a set of commuting analytic difference operators with coefficients being elliptic functions. Often these correspond to well known relativistic elliptic integrable models (and sometimes more esoteric ones). Thus we will refer to the set of ${\cal H}_\alpha$ as an elliptic integrable model. See \cite{Gaiotto:2012xa,Gaiotto:2015usa,Razamat:2018zel,Nazzal:2018brc,Nazzal:2021tiu,Nazzal:2023bzu} for concrete examples.

\

There are various ways to derive the integrable models associated to the $6d$ theory (with a given circle compactification). One such way is first to derive, or more precisely conjecture, an across dimensions dual of some compactification with enough punctures. This means to conjecture a $4d$ Lagrangian theory flowing in the IR to the same fixed point as the $6d$ SCFT compactified on a surface. Then using the Lagrangian theory one can compute the corresponding index and derive the integrable model from its analytical properties \cite{Gaiotto:2012xa}.\footnote{See \cite{Razamat:2013jxa,Gaiotto:2015usa,Maruyoshi:2016caf,Ito:2016fpl,Yagi:2017hmj} for related developments.} 

\subsection{Index and eigenfunctions}

Given the Kernel property of the index \eqref{Eq:kernel} it is natural to wonder whether one can expand the index in terms of some proper set $\Lambda$ of eigenfunctions of the integrable model,
\be 
{\cal H}_\alpha\left[\textcolor{blue}{T_{6d},G_{5d}}\right]({\bf x},{\bf u}_{6d};q,p)\cdot \psi_\lambda({\bf x}) =
E_{\alpha,\lambda}\, \psi_\lambda({\bf x})\,. 
\ee Because of self-adjointnes  of the Hamiltonians we can choose these functions to form an orthonormal set,
\be 
\oint d{\bf x} \;\Delta({\bf x}, {\bf u}_{6d};q,p)\;\psi_\lambda({\bf x})\;\psi_{\lambda'}({\bf x}^{-1}) = \delta_{\lambda,\lambda'}\,.
\ee 

We want to make the following ansatz,
\be 
{\cal I}[\textcolor{blue}{T_{6d}},\textcolor{red}{\cal C}](\{{\bf x}_j\},{\bf u}_{6d},q,p)=
\sum_{\lambda\in \Lambda} C_\lambda[\textcolor{blue}{T_{6d}},\textcolor{red}{\cal C}]({\bf u}_{6d};q,p)
\prod_{j=1}^s\,\psi_{\lambda}({\bf x}_j)\,.
\ee Here we only refine the index with fugacities corresponding to the chosen type of a maximal puncture (and $s$ is the number of such punctures). If such an ansatz makes sense then the Kernel property is manifest. Moreover, if we glue two surfaces together then the index of the glued surface is given by,
\be\label{Eq:ansatz}
&&{\cal I}[\textcolor{blue}{T_{6d}},\textcolor{red}{{\cal C}_1\oplus {\cal C}_2}](\,\{{\bf x}_j\}_{S_1\cup S_2},{\bf u}_{6d},q,p)=\\
&&\qquad\qquad\qquad\sum_{\lambda\in \Lambda} C^{(1)}_\lambda[\textcolor{blue}{T_{6d}},\textcolor{red}{{\cal C}_1}]({\bf u}_{6d};q,p)\;
C^{(2)}_\lambda[\textcolor{blue}{T_{6d}},\textcolor{red}{{\cal C}_2}]({\bf u}_{6d};q,p)
\prod_{j\in S_1\cup S_2}\,\psi_{\lambda}({\bf x}_j)\,.\nonumber 
\ee One way for such an expansion to make sense is if there is a natural ordering on $\lambda\in \Lambda$,
\be \label{Eq:ordering}
\lambda_0 \leq \lambda_1 \leq \lambda_2\leq \lambda_3\leq \cdots .
\ee We will compute the indices in expansion in the parameters $q$ and $p$ (assuming as before that these are taken to be inside the unit circle). 
We note that the index of a superconformal theory is always regular in such an expansion and starts off as ${\cal I}=1+\cdots$.
We define partial sums,
\be
{\cal I}^{(n)} =\sum_{i=0}^n C_{\lambda_i}[\textcolor{blue}{T_{6d}},\textcolor{red}{\cal C}]({\bf u}_{6d};q,p)
\prod_{j=1}^s\,\psi_{\lambda_i}({\bf x}_j)\,.
\ee If we want to reproduce the index up to any given order $N$ in expansion in $q$ and $p$ and there is a finite value of $n(N)$ such that up 
to order $N$, ${\cal I}^{(n>n(N))}$ are equal, then the ansatz makes sense. This amounts to the coefficients $C_{\lambda_i}$ 
contributing at non-decreasing orders as we increase $i$. We will give evidence in several examples below that this property is in fact true for a wide variety of setups.

Moreover, we will see that the eigenfunction $\psi_{\lambda_0}({\bf x})$ play a special role.  We will find that there is a unique eigenfunction
which contributes to the index at order $N=0$. Thus, in particular $\lambda_0$ is strictly less than $\lambda_i$ for all $i>0$. We will refer to this 
eigenfunction as the ground state of the integrable system and will denote it by $\psi_0({\bf x})$.

\begin{figure}[htbp]
	\centering
  	\includegraphics[scale=0.4]{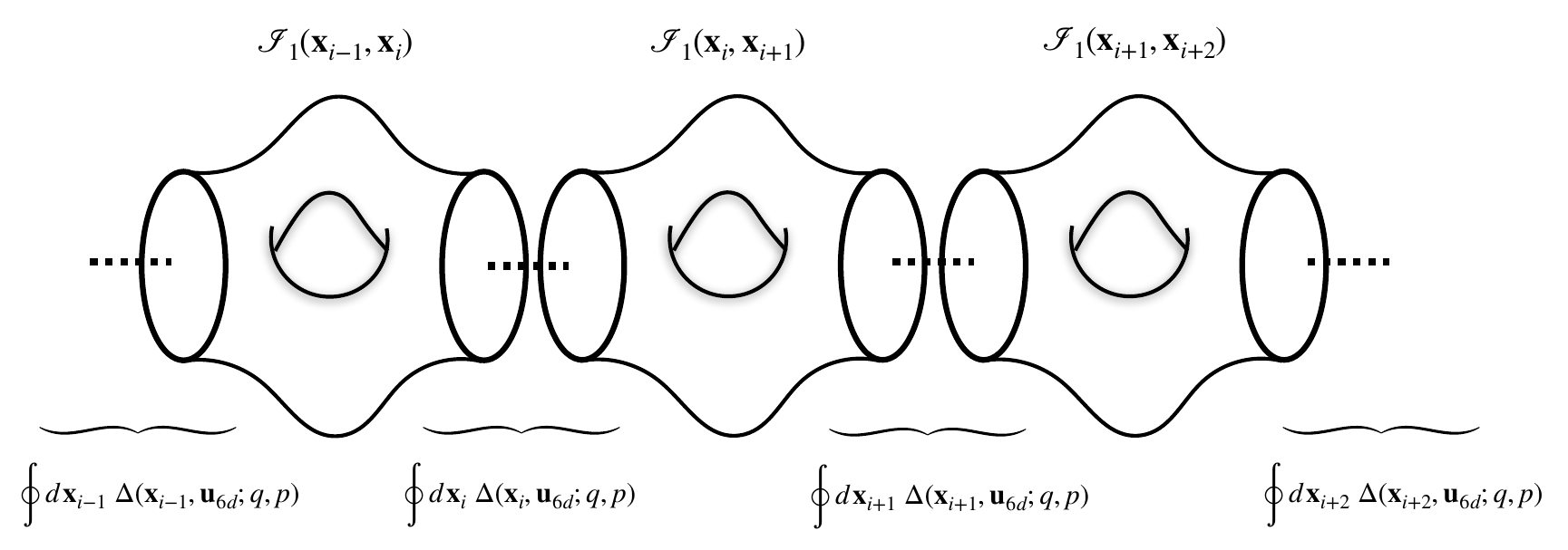}
    \caption{Graphical representation of gluing indices of tori with (at least) two maximal punctures to higher genus surfaces.
    }
    \label{F:DualLattices}
\end{figure}

\subsection{Ground state from large compactifications}

Let us assume that we can compute the index using explicit Lagrangian across dimension dualities for some compactification with at least two maximal punctures. In addition we might have other punctures, the surface might have some higher genus, and there might be some flux. We assume that the ansatz for the index of this theory in terms of eigenfunctions \eqref{Eq:ansatz} is well defined in the sense discussed above. Then we write the index for this theory as,
\be 
{\cal I}_1({\bf x}_1,{\bf x}_2) = \sum_{i=0}^\infty C_{\lambda_i}\; \psi_{\lambda_i}({\bf x}_1)\; \psi_{\lambda_i}({\bf x}_2)\,.
\ee Next we consider the index obtained by gluing $\textcolor{blue}{n}$ copies of this theory sequentially along the maximal punctures. This can be done iteratively,
\be
{\cal I}_{n+1}({\bf x}_1,{\bf x}_2)= \oint d{\bf x} \;\Delta({\bf x}, {\bf u}_{6d};q,p)\; {\cal I}_n({\bf x}_1,{\bf x})\; {\cal I}_1({\bf x}^{-1},{\bf x}_2)\,.
\ee The theory obtained in this way has the number of other punctures, the genus, and the flux, multiplied by $\textcolor{blue}{n}$. The index is given in terms of eigenfiunctions by,
\be 
{\cal I}_{\textcolor{blue}{n}} ({\bf x}_1,{\bf x}_2) = \sum_{i=0}^\infty \left(C_{\lambda_i}\right)^{\textcolor{blue}{n}}\; \psi_{\lambda_i}({\bf x}_1)\; \psi_{\lambda_i}({\bf x}_2)\,.
\ee We then take the limit of large $\textcolor{blue}{n}$. Up to any set order of the expansion of the index in this limit, starting with some value of $\textcolor{blue}{n}$ only the ground state will contribute to the index. In paticular we can compute,
\be \label{Eq:C0}
C_{0}\equiv C_{\lambda_0}= \lim_{\textcolor{blue}{n}\to \infty} \frac{{\cal I}_{\textcolor{blue}{n+1}} ({\bf x}_1,{\bf x}_2)}{{\cal I}_{\textcolor{blue}{n}} ({\bf x}_1,{\bf x}_2)}\,.
\ee From here we obtain an explicit expression for the ground state eigenfunction,
\be \label{Eq:psi0}
\widetilde \psi_0({\bf x})\equiv \psi_0({\bf 1})\; \psi_0({\bf x})= \lim_{\textcolor{blue}{n}\to \infty}\frac{1}{\left(C_0\right)^{\textcolor{blue}{n}}}\,
{\cal I}_{\textcolor{blue}{n}} ({\bf x},1)\,.
\ee  Here we can get rid of $\psi_0({\bf 1})$ factor by normalizing the eigenfunction using the appropriate scalar product. This gives 
us a very simple algorithm to compute the ground state eigenfunctions for a variety of elliptic relativistic integrable models. The algorithm only relies on the physical input  of deriving across dimensions dualities for a two punctured compactification on some surface satisfying the relevant properties outlined above. In principle one can try and generalize the algorithm to derive all eigenfunctions of a given model. However the details will cease to be generic as we might  have several $\lambda_i$ contributing at the same order of the expansion. Moreover, it will be technically more involved to implement as it would require computation of indices to high orders in expansion to derive eigenfunctions up to   low orders.

\

In the following section we will implement this algorithm to derive ground state eigenfunctions for a variety of models.

\subsection{Universality of large compactifications}
\label{sec:universality}

Let us next discuss the physical implication of the above. For concreteness let us take the theory with two punctures to be a genus one compactification with two maximal punctures and no flux. Then gluing $\textcolor{blue}{g-1}$ such theories together we will obtain a theory 
corresponding to genus $\textcolor{blue}{g-1}$  compactification and two maximal punctures. Finally we can glue the two punctures together to obtain a theory corresponding to genus $\textcolor{blue}{g}$ compactification with no flux and no punctures. Following the above results the index of this theory in the large $\textcolor{blue}{g}$ limit is well approximated by,
\be
{\cal I}_{\textcolor{blue}{g}} \sim \left(C_0\right)^{\textcolor{blue}{g-1}}\,.
\ee By well approximated we mean that the deviation between two sides of the above starts at orders in expansion which grow linearly with $\textcolor{blue}{g}$. This universal result hints that there should be a clean physical interpretation associated with $C_0$. In fact it was conjectured in \cite{GaiottoEtAl} that the expansion \eqref{Eq:ansatz} has the following meaning. Let us again take for concerteness the case say of genus $g$ compactification with no flux and no punctures so that \eqref{Eq:ansatz} takes the form of,
\be
{\cal I}= \sum_{\lambda\in \Lambda} \left(C_\lambda\right)^{\textcolor{blue}{g-1}}\,.
\ee The index counts (with signs) various local operators in the $4d$ theory. The $4d$ theory is a compactification of a $6d$ one. A natural question is whether one can identify the origin of the $4d$ operators counted by the index in the $6d$ theory. The conjecture of \cite{GaiottoEtAl} is that the $C_0$ captures local operators in $4d$ which originate from local operators in $6d$ properly smeared on the Riemann surface. The smearing (which is the essence of the Riemann-Roch theorem) resonates with the $g-1$ power appearing in the index. For other values of $\lambda$ the local operators in $4d$ originate from non-local operators in $6d$ wrapping the Riemann surface. The label $\lambda$ should be related to labeling of various such operators in $6d$. In the limit of large genus the fact that the index is dominated by $\lambda=\lambda_0$ indicates then that  the non-local operators would acquire large charges upon compactification and contribute to the tail of the expansion of the index. 

The function $C_0$ captures directly  information about local operators of the $6d$ theory. Technically one can view it as contribution to the index in $4d$ of local operators in $6d$ in genus two compactifications. At the more conceptual level the coefficients of various terms in $C_0$ count with signs dimensions of certain vector spaces one can associate to local operators in $6d$ \cite{babuip}. These dimensions depend on the quantum numbers of the theory, and in particular also on ones related to the compactified geometry.  Note also that here as we do not have flux and because of the universality in large genus limit the coefficient $C_0$ should manifest the full symmetry $G_{6d}$. This should be also true for higher coefficients $C_\lambda$. As the coefficient $C_0$ for compactifications with zero flux and no punctures will play an important role in understanding the $6d$ physics, to distinguish it from other compactifications we will denote it by $\hat C_0$.
By a similar logic \cite{GaiottoEtAl} the eigenfunction $\psi_0({\bf x})$ should capture the circle reduction of $5d$ operators associated to the  maximal puncture. 

\section{Examples}\label{sec:examples}

\subsection{Elliptic RS model}

The first example we will discuss is the elliptic Ruijsenaars-Schneider model of type $\A_1$. The action of the basic Hamiltonian on a function is given by,\footnote{Note that our choice of parameters is slightly different than the usual one in the literature. For example our $t$ is related to the one in \cite{Gaiotto:2012xa} as $t\to t \left(q\,p\right)^{-\frac12}$. }
\be
{\cal H}^{RS}_{\A_1}\cdot \psi(x) = \frac{\th(\sqrt{\frac{p}{q}}\,t\,x^{-2})}{\th(x^2)}\,\psi(x \,q^{\frac12})+
 \frac{\th(\sqrt{\frac{p}{q}}\,t\,x^{2})}{\th(x^{-2})}\,\psi(x \,q^{-\frac12})\,.
\ee This model arises when compatifying the $(2,0)$ type $\A_1$ $6d$ SCFT to $4d$ \cite{Gaiotto:2012xa}.
The symmetry of the $6d$ theory is $G_{6d}=\SU(2)$, fugacity for the Cartan generator of which is parametrized here by $t$. The circle compactification to $5d$ gives the maximally supersymmetric YM theory with gauge group $\SU(2)$. Thus the maximal puncture symmetry is $\SU(2)$ and the fugacity for it  is the parameter (denoted by $x$ above) on which the Hamiltonian acts.

The simple across dimension duality we can use is the compactification on a sphere with two maximal $\SU(2)$ punctures (and a third $\SU(2)$ puncture that will play no role). The relevant theory is just a collection of two free bifundamental chiral fields \cite{Gaiotto:2009we}. The index is given by,
\be 
{\cal I}_1({\bf x}^1,{\bf x}^2) =  \G\left((q\,p)^{\frac14}t^{\frac12} (x^1)^{\pm1} (x^2)^{\pm1}\right)^2\,.
\ee  The integration measure here is,
\be 
\oint d{\bf x} \;\Delta(x, {\bf u}_{6d};q,p)\; \cdots =\frac{(q;q)(p;p)}2 \oint \frac{dx}{2\pi i x}
\frac{\G(\frac{\sqrt{q\,p}}{t}\,x^{2})\,\G(\frac{\sqrt{q\,p}}{t}\,x^{-2})\,\G(\frac{\sqrt{q\,p}}{t})}{\G(x^{2})\G(x^{-2})}\cdots\,.
\ee 
We use the following definitions,
\be
\G(z)=\prod_{i,j=0}^\infty\frac{1-z^{-1}q^{i+1}p^{j+1}}{1-z\, q^ip^j}\,,\; \th(z)=\prod_{i=0}^\infty (1-z\, p^i)(1-z^{-1}p^{i+1})\,,\; (z;q)=\prod_{i=0}^\infty (1-z\,q^i)\,.\;\;\;\;\;\;
\ee
Using these definitions  we readily compute from \eqref{Eq:C0},
\be
C_0 =1+3\,t\,\sqrt{p\,q}+\left(t^{-2}-2+5t^2\right)p\,q+\left(3t-t^{-1}\right)(p^{\frac32}\,q^{\frac12}+q^{\frac32}\,p^{\frac12})+\cdots\,.
\ee Our procedure gives results as an expansion in parameters $q$ and $p$. We will quote the results to the orders we were able to perform actual computations. However, in principle one can compute to any desired order.\footnote{Practically, one can take $q/p\equiv y$ and $\sqrt{qp}\equiv X$ and think of all our expressions as expansions in $X$. The $\cdots$ in the expressions denote higher orders in $X$. Interested reader can consult a Mathematica notebook (\href{https://github.com/anedelin/GroundStates}{https://github.com/anedelin/GroundStates}) for details of the  computation.}

Using \eqref{Eq:psi0} we then obtain,
\be \label{Eq:RSeig}
&&\widetilde \psi_0(x) = 1+(4+x^2+x^{-2})\, t\,(q\,p)^{\frac12}+ \left(t^{-1}+(5+x^2+x^{-2})\,t\right)(q+p)\,(q\,p)^{\frac12}+\\
&&\;\;\; \left(10\,t^2-6-x^2-x^{-2}-t^{-2}+t^2(x^4+x^{-4}+4x^2+4x^{-2})\right)q\,p+\cdots\,.\nonumber
\ee 
We verify (in expansion in $q$ and $p$ up to an order we could perform the computation) that the above is an eigenfunction,
\be 
{\cal H}^{RS}_{\A_1}\cdot \widetilde \psi_0(x) =E_0 \, \widetilde \psi_0(x)\,,
\ee 
and obtain that the ground state energy is,
\be
E_0 = 1-p+(t+\frac1t)\,\sqrt{p\,q}-pq+(t+\frac1t)\,p\sqrt{p\,q}-p^2+\cdots\,.
\ee
Note that $t$ parametrizes the Cartan generator of the $G_{6d}=\SU(2)$ and the energy is invariant under the Weyl group of this symmetry. 
An additional identity $\widetilde \psi_0(x)$ has to satisfy \cite{Gaiotto:2012xa} is,
\be
\left.\widetilde \psi_0(x)\right|_{t\to\frac1t} =\G\left(\frac{\sqrt{q\,p}}{t}\,x^{2}\right)\,\G\left(\frac{\sqrt{q\,p}}{t}\,x^{-2}\right)\,\G\left(\frac{\sqrt{q\,p}}{t}\right)^4 \, \widetilde \psi_0(x)\,,
\ee can be also verified to hold for the eigenfunction \eqref{Eq:RSeig}  given here. 

We also can consider various limits of the index giving simple known eigenfunctions \cite{Gadde:2011uv,Gadde:2011ik}. First let us consider the Schur limit. In our notations this corresponds to taking $t=\left(q/p\right)^{\frac12}$. Then the eigenfunctions $\psi_\lambda(x)$ are just the Schur polynomials times \\$1/\left((q\,x^2;q)(q\,x^{-2};q)(q;q)\right)$. This can be easily verified to hold for  \eqref{Eq:RSeig}. Moreover taking first $t\to t \left(q\,p\right)^{-\frac12}$ and then $p\to 0$ we obtain the Macdonald limit of the index with the eigenfunctions expected to be given by Macdonald polynomials times $1/\left((t\,x^2;q)(t\,x^{-2};q)(t;q)\right)$. Again, this can be verified to hold for \eqref{Eq:RSeig}. The relevant polynomial in both cases is just the constant one. More explicitly,
\be 
\left.\widetilde \psi_0(x)\right|_{t\to t \left(q\,p\right)^{-\frac12}|p\to 0} =\frac{(t^2;q)}{(t\,q;q)}\frac1{(t\,x^2;q)(t\,x^{-2};q)(t;q)^4}\,.
\ee  Finally, in this case by studying the Macdonald limit we know that the labels $\lambda$ in \eqref{Eq:ansatz} correspond to finite dimensional irreps of $\SU(2)$ \cite{Gadde:2011uv}. From here we deduce that there should be no  degeneracy for $\lambda_i$, and thus \eqref{Eq:ordering} should define a strict ordering. From here we define
\be \label{Eq:C1}
C_{1}\equiv C_{\lambda_1}= \lim_{\textcolor{blue}{n}\to \infty} \frac{{\cal I}_{\textcolor{blue}{n+1}}(x,1)-(C_0)^{\textcolor{blue}{n+1}}\widetilde \psi_0(x)} {{\cal I}_{\textcolor{blue}{n}} (x,1)-(C_0)^{\textcolor{blue}{n}}\widetilde \psi_0(x)}\,.
\ee The explicit computation gives,
\be 
C_1= 2(q\,p)^{\frac14}\,\left(\sqrt{t}-3 p q \sqrt{t}+(3p+3q+2)\sqrt{pq}  t^{3/2}+3 p q t^{5/2}+\cdots\right)\,.
\ee Note that $C_1$ leading term scales as $(q\,p)^{\frac14}$ whereas for $C_0$ it scales as $1$ and that is the reason why we can separate the two contributions in the limit considered here. From here we obtain an expression for
the first excited state eigenfunction,
\be \label{Eq:psi1} 
\widetilde \psi_1({\bf x})\equiv \psi_1({\bf 1})\; \psi_1({\bf x})= \lim_{\textcolor{blue}{n}\to \infty}\frac{1}{\left(C_1\right)^{\textcolor{blue}{n}}}\,
\left({\cal I}_{\textcolor{blue}{n}} ({\bf x},1)-(C_0)^{\textcolor{blue}{n}}\widetilde \psi_0(x)\right)\,.
\ee The computation results in,
\be
\widetilde \psi_1(x) = 2\,(x+\frac1x)\left(1+p+q-(\frac{1}{t}-4
    t)\,\sqrt{p\,q}\right) +2\,(x^3+\frac1{x^3})\,t\, \sqrt{q\, p}+\cdots\,.
\ee In principle we can continue to other eigenfunctions in a similar manner. 
We can also compute $\hat{C}_0$ coefficient which in this case is given by:
\be
\hat{C}_0=1+pq\left(t^2+t^{-2}+4\right)-2\sqrt{pq}(p+q)\left(t+t^{-1}\right)+pq(p+q)\left(t^2+5+t^{-2}\right)
\nonumber\nn\\
-4(pq)^{3/2}\left(t+t^{-1}\right)-2\sqrt{pq}(p^2+q^2)\left(t+t^{-1}\right)+\,\cdots\,.
\label{C0hat:RS}
\ee Note that the term at order $q\,p$ can be written as $\chi_{adj.\,\SU(2)}(t)+3$. This looks as a contribution of a conserved current of global $\SU(2)_t$ symmetry of the $6d$ $(1,0)$ theory and we might want to interpret the $+3$ as coming from additional rotations in the compactification dimensions. This is very reminiscent of indices of compactifications on a sphere \cite{Hwang:2021xyw} where the rotations, the $\SU(2)$ isometry of the sphere, becomes a global symetry in $4d$.

\subsection{The $\A_2$  model}

Let us start with the following Hamiltonian defined on $\A_2$ root system,
\be \label{Eq:A2Hamiltonian}
&&{\cal H}^{\A_2}_Y\cdot \psi({\bf x}) = \frac{\th(p^\frac12 Y x_2/x_3)\th(p^\frac12 Y x_3/x_2)}{\th(x_2/x_1)\th(x_3/x_1)}\psi(x_1 q^{-\frac23},x_2 q^{\frac13},x_3 q^{\frac13})+\nonumber\\
&&\qquad\qquad\frac{\th(p^\frac12 Y x_1/x_3)\th(p^\frac12 Y x_3/x_1)}{\th(x_1/x_2)\th(x_3/x_2)}\psi(x_1 q^{\frac13},x_2 q^{-\frac23},x_3 q^{\frac13})
 +\\ 
&& \frac{\th(p^\frac12 Y x_2/x_1)\th(p^\frac12 Y x_1/x_2)}{\th(x_1/x_3)\th(x_2/x_3)}\psi(x_1 q^{\frac13},x_2 q^{\frac13},x_3 q^{-\frac23})\,.\nonumber
\ee Here $Y$ is a general parameter and Hamiltonians with different $Y$ commute with each other.
The parameters $x_i$ satisfy $\prod_{i=1}^3x_i=1$. This Hamiltonian was derived in \cite{Razamat:2018zel} as corresponding to the integrable system associated with the $6d$ SCFT being the so called minimal $\SU(3)$ SCFT \cite{Seiberg:1996qx,Bershadsky:1997sb} and the $5d$ effective theory being pure Chern-Simons $\SU(3)$ model with level nine \cite{Jefferson:2018irk}. The relevant across dimensions duality was derived in \cite{Razamat:2018gro}. 
The model was further discussed in \cite{Ruijsenaars:2020shk}.

The simple across dimension duality we can use is the compactification on a sphere with two maximal $\SU(3)$ punctures (and two so called empty ones). The relevant theory is just a collection of three free bifundamental chiral fields with a baryonic superpotential \cite{Razamat:2018gro}. The index is given by,
\be 
{\cal I}_1({\bf x}^1,{\bf x}^2) = \prod_{i,j=1}^3 \G\left((q\,p)^{\frac13} x^1_i x^2_j\right)^3\,.
\ee  The integration measure here is,
\be 
\oint d{\bf x} \;\Delta({\bf x}, {\bf u}_{6d};q,p)\; \cdots =\frac{(q;q)^2(p;p)^2}6 \oint \prod_{i=1}^2\frac{dx_i}{2\pi i x_i}
\prod_{i=1}^3\prod_{j=i+1}^3\frac1{\G(x_i/x_j)\G(x_j/x_i)}\cdots\,.
\ee Note here $G_{6d}$ is trivial and thus there are no ${\bf u}_{6d}$ parameters. 

Using these ingredients we readily compute from \eqref{Eq:C0},
\be
&&C_0=1+2 p q+2 p^2 q+3 p^3 q+3 p^4 q+3 p^5 q+3 p^6 q+2 p q^2+6 p^2 q^2+7 p^3 q^2+\nn\\
&&9 p^4 q^2+11p^5 q^2+3 p q^3+7 p^2 q^3+9 p^3 q^3+13 p^4 q^3+3 p q^4+9 p^2 q^4+ 13 p^3 q^4+\nn\\
&&\hspace{7cm} 3 p q^5+11 p^2 q^5+3 p q^6+\cdots \,.
\ee Using \eqref{Eq:psi0} we then obtain the ground state eigenfunction,
\be
&&\widetilde \psi_0({\bf x}) =1+\left(\frac{1}{x_1^3}+\frac{1}{x_2^3}+\frac{1}{x_3^3}\right)\,q\,p +\left(\frac{1}{x_1^3}+\frac{1}{x_2^3}+\frac{1}{x_3^3}-1\right)(q^2p+p^2q)+\\
&&\;\; \left(\frac{1}{x_1^6}+\frac{1}{x_2^6}+\frac1{x_3^6}+\frac{1}{x_1^3}+\frac{1}{x_2^3}+\frac1{x_3^3}+x_1^3+x_2^3+x_3^3-4\right) \,q^2p^2+\nn\\
&&\left(\frac{1}{x_1^3}+\frac{1}{x_2^3}+\frac{1}{x_3^3}-3\right) (q^3p+p^3q)+\left(\frac{1}{x_1^3}+\frac{1}{x_2^3}+\frac{1}{x_3^3}-3\right)(q^4p+p^4q)+\nonumber\\
&& \left(\frac{1}{x_1^6}+\frac{1}{x_2^6}+\frac1{x_3^6}+x_1^3+x_2^3+x_3^3-9\right)(q^3p^2+p^3q^2)+\cdots \,. \nonumber
\ee It can be verified that indeed $\widetilde \psi_0({\bf x})$ is an eigenfunctions of \eqref{Eq:A2Hamiltonian}, 
\be
{\cal H}^{\A_2}_Y\cdot \widetilde \psi_0({\bf x}) =E_0 \, \widetilde \psi_0({\bf x})\,,
\ee  and the corresponding ground state energy $E_0$ is given by,
\be 
E_0 = \frac{(1-p^{\frac32} Y^3)(1-p^{\frac32} Y^{-3})}{\th(p^{\frac12}Y)}(1-q\,p -p^3-p\,q^2-2q\,p^2-3p^3q-2p^2q^2+\cdots)\,.
\ee This expression was verified up to fourth order in the expansion in $p$ and $q$.     
Finally, we can also compute the coefficient $\hat C_0$ which is given by,
\be
&&\hat C_0 =\\
&&\;1+3 q\,p+4(q^2p+p^2q)+16(q\,p)^2+9(q^3p+p^3q)+30(q^3p^2+p^3q^2)+9(q^4p+p^4q)+\cdots\,.\nonumber
\ee Note here at the order $q\,p$ we only have the ``geometric'' $+3$ coefficient as there is no global symmetry in $6d$.

\

\subsection{The $\A_3$ model}

Now let's move to the $A_3$ model and repeat the same procedure. The Hamiltonian for this model is defined by,

\be \label{Eq:A3Hamiltonian}
&&{\cal H}^{\A_3}\cdot \psi({\bf x}) = \frac{\th(p^\frac12  x_3/x_2)\th(p^\frac12  x_4/x_2)\th(p^\frac12  x_3/x_4)}{\th(x_2/x_1)\th(x_3/x_1)\th( x_4/x_1)}\psi(x_1 q^{-\frac34},x_2 q^{\frac14},x_3 q^{\frac14},x_4 q^{\frac14})+\nonumber\\
&&\qquad\qquad\frac{\th(p^\frac12  x_3/x_1)\th(p^\frac12  x_1/x_4)\th(p^\frac12  x_3/x_4)}{\th(x_1/x_2)\th(x_3/x_2)\th( x_4/x_2)}\psi(x_1 q^{\frac14},x_2 q^{-\frac34},x_3 q^{\frac14},x_4 q^{\frac14})
+
\nonumber\\
&&\qquad\qquad\frac{\th(p^\frac12  x_1/x_2)\th(p^\frac12  x_4/x_2)\th(p^\frac12  x_1/x_4)}{\th(x_2/x_3)\th(x_1/x_3)\th( x_4/x_3)}\psi(x_1 q^{\frac14},x_2 q^{\frac14},x_3 q^{-\frac34},x_4 q^{\frac14})
+\\ 
&&\qquad\qquad\frac{\th(p^\frac12  x_3/x_2)\th(p^\frac12  x_1/x_2)\th(p^\frac12  x_3/x_1)}{\th(x_3/x_4)\th(x_2/x_4)\th( x_1/x_4)}\psi(x_1 q^{\frac14},x_2 q^{\frac14},x_3 q^{\frac14},x_4 q^{-\frac34})\,.\nonumber
\ee
where $\prod_{i=1}^4 x_i=1$ parametrize the $\SU(4)$ puncture symmetry on which the operator acts. Note that as opposed to the $A_2$ case there are no additional parameters that the operator depends on. This operator was derived in \cite{Razamat:2018zel} from compactifications of $\SO(8)$ minimal conformal matter and was also discussed later in \cite{Ruijsenaars:2020shk}.

Similar to the previous case, the simplest across dimension duality we have here is the compactification on a sphere with two maximal $\SU(4)$ punctures and two empty ones ({\it i.e.} the puncture with no symmetry). The relevant theory consists of just two bifundamental chiral fields and a baryonic superpotential \cite{Razamat:2018gro}. The superconformal index is given by, 
\be 
{\cal I}_1({\bf x}^1,{\bf x}^2) = \prod_{i,j=1}^4 \G\left((q\,p)^{\frac14} x^1_i x^2_j\right)^2\,.
\ee  The integration measure here is,
\be 
\oint d{\bf x} \;\Delta({\bf x}, {\bf u}_{6d};q,p)\; \cdots =\frac{(q;q)^3(p;p)^3}{24} \oint \prod_{i=1}^3\frac{dx_i}{2\pi i x_i}
\prod_{i=1}^4\prod_{j=i+1}^4\frac1{\G(x_i/x_j)\G(x_j/x_i)}\cdots\,.\;
\ee Also here $G_{6d}$ is trivial and thus there are no ${\bf u}_{6d}$ parameters. 
\\
Performing the same computation as above we find from \eqref{Eq:C0},

\be
C_0=1+2 p q + 2 p^2 q + 2 p^3 q + 2 p q^2 + 4 p^2 q^2 + 2 p q^3+\cdots
\ee
Then using \eqref{Eq:psi0} we find the ground state to be,

\be
&&\widetilde \psi_0({\bf x}) =1+\left( 3+\sum_{i<j}^4 x_i^2x_j^2 \right)pq+
\left( 2+\sum_{i<j}^4 x_i^2x_j^2 \right) (pq^2 +p^2q +pq^3+p^3 q)\nonumber
+\\
&&\hspace{4cm}\;\; \left(\sum_{i<j}^4 x_i^4x_j^4+ 5\sum_{i<j}^4 x_i^2x_j^2  +\sum_{i<j}^4 x_i^{-4}\right)p^2 q^2
+\cdots \, 
\ee
We can act with \eqref{Eq:A3Hamiltonian} to verify that this is an eigenfunction of the Hamiltonian. Indeed we find that the energy is given by,
\be
E_0=1 + 2 p^\frac12 + 2 p + 4 p^\frac32 + 5 p^2 - p q+\cdots
\ee

\subsection{The van Diejen model}

Another interesting example of elliptic integrable Hamiltonians is van Diejen integrable model that was first introduced as deformation of RS model in 
\cite{vanDiejen}.\footnote{See {\it e.g.} \cite{MR2299829,MR3313680,MR4521699} for discussions of some of the eigenfunctions of this model.} The Hamiltonian itself is written as follows \cite{MR4120359},
\be
{\cal H}^{vD}\cdot \psi(x) \equiv \frac{\prod\limits_{n=1}^8\theta_p\big((pq)^{\frac12}h_n x\big)}{\theta_p(x^2)\theta_p\big(qx^2\big)}\;\psi(qx)+\frac{\prod\limits_{n=1}^8\theta_p\big((pq)^{\frac12}h_n x^{-1}\big)}{\theta_p(x^{-2})\theta_p\big(qx^{-2}\big)}\;\psi\big(q^{-1}x\big)
+V(h;x)\;\psi(x)\,.\nonumber\\
\label{vanDiejen:hamiltonian}
\ee
\noindent This operator depends on the octet of $h_i$ parameters. 
The constant term $V(h;x)$ of van Diejen Hamiltonian is an elliptic function in $x$ variable with periods $1$ and $p$. Poles of 
this function in the fundamental domain are located at 
\be
x=\pm q^{\pm\frac{1}{2}}\,,\qquad x=\pm q^{\pm \frac{1}{2}}p^{\frac{1}{2}}\,,
\label{Diejen:poles}
\ee
and corresponding residues are given by:
\be
\mathrm{Res}_{x=sq^{\pm 1/2}}V(h;x)=\mp s\frac{\prod\limits_{n=1}^8\thf{sp^{\frac{1}{2}}h_n}}{2q^{\mp \frac{1}{2}}\thf{q^{-1}}\qPoc{p}{p}^2}\,,\nn\\
\mathrm{Res}_{x=sq^{\pm 1/2}p^{1/2}}V(h;x)=\mp s\frac{\prod\limits_{n=1}^8h_n^{-\frac{1}{2}}\thf{s h_n}}{2q^{\mp\frac{1}{2}}p^{-\frac{3}{2}}\thf{q^{-1}}\qPoc{p}{p}^2}\,,
\label{Diejen:residues}
\ee
where $s=\pm1$. The expression we can write for this constant term is not unique. For our purposes we will use the following form,
 \be
&&V(x;h_i)=
 \frac{\prod\limits_{j\neq i}^{8}\thf{q^{-1} h_{i}h_jh^{-\frac{1}{2}}}}
 {\thf{q^{-2}h_{i}^{2}h^{-1/2}}}
\frac{\thf{\pq{1}{2}h_{i}^{-1}x^{\pm1}}}{\thf{\pqm{1}{2}h_{i}h^{-1/2}q^{-1}x^{\pm 1}}}+
\nn\\
&&
\left[\frac{\prod\limits_{j\neq i}^{8}\thf{\pq{1}{2}h_jx^{-1}}}{\thf{\pq{1}{2}h_{i}^{-1}h^{1/2}q x^{-1}}}
\frac{\thf{\pq{1}{2}h_{i}^{-1}h^{1/2}x}\thf{\pq{1}{2}h_{i}^{-1}x^{-1}}}
{\thf{q^{-1}x^{-2}}\thf{x^2}}+\left( x\to x^{-1} \right)\right]\,,
\label{a1:op:noflip:gen}
\ee
which was derived in \cite{Nazzal:2021tiu}. Here $h$ parameter is just a product of all $h_i$:
\be
h=\prod_{i=1}^8 h_i\,.
\ee
This model was shown to arise in the compactifications of the $6d$ E-string theory down to $4d$ \cite{Nazzal:2018brc,Nazzal:2021tiu}.  
The global symmetry of $6d$ theory here is ${\text E}_8$ whose Cartan is parametrized by $h_i$ parameters. Circle compactification of the E-string theory 
leads to the $\SU(2)$ maximally supersymmetric Yang-Mills theory. Thus the maximal puncture symmetry in $4d$ theory is also $\SU(2)$ and van Diejen operator 
\eqref{vanDiejen:hamiltonian} acts on the fugacities of this symmetry. 
From the point of view of $4d$ theory $h_i$ parameters play the role of the inverse charges of the $\SU(2)$ puncture moment maps. 
The moment maps depend on the type of the puncture. In what follows for convenience
reasons we will use such a puncture  that all of the parameters of the integrable models
above are identified with moment maps directly.
Considering the  problem with all parameters turned on is computationally
complicated so we will only analyze it  by setting all of the moment maps 
charges to be the same and equal to $t$, which at  the level of $h_i$ parameters is equivalent to,
\be
h_i=t^{-1}\,,~\forall~~ i=1,...,8\,.
\label{vD:parameters}
\ee

The simplest building blocks we need to run our arguments are two- and three-punctured spheres with $\SU(2)$ punctures. Two-punctured sphere (tube) 
theory is just a bifundamental chiral field for two $\SU(2)$ symmetries with the flip field and two octets of the fundamental multiplets corresponding to 
two sets of the moment map operators \cite{Gaiotto:2015una,Kim:2017toz}. In case all moment maps have the same charges the index reads, 
\be
{\cal I}_1(x_1,x_2)=\GF{pqt^{4}}\GF{t^{-2}x_1^{\pm 1}x_2^{\pm 1}}\GF{\pq{1}{2}t x_1^{\pm 1}}^{8}\,.
\label{vD:tube}
\ee
Here we define tube theory so that the punctures are of conjugated types. The integration measure for the resulting eigenfunctions is the following:
\be
\oint d{\bf x} \;\Delta(x, {\bf u}_{6d};q,p)\; \cdots =\frac{(q;q)(p;p)}2 \oint \frac{dx}{2\pi i x}\frac{1}{\GF{x^{\pm 2}}}
\prod\limits_{i=1}^8\GF{\pq{1}{2}h_i x^{\pm 1}}\cdots\,=
\nn\\
\frac{(q;q)(p;p)}2 \oint \frac{dx}{2\pi i x}
\frac{1}{\GF{x^{\pm 2}}}\GF{\pq{1}{2} t^{-1}x^{\pm 1}}^8\cdots\,,
\ee
where in the last line we specify measure in our case corresponding to \eqref{vD:parameters}.
Then using our algorithm we derive the following expression for $C_0$:
\be
C_0=1-pq(1+p+q)t^{-4}+\left(t^4+28\left(t^2-t^{-2}\right)\right)pq(1+p+q+p^2+q^2)+
\nn\\p^2 q^2 \left(t^8+28 t^6+273 t^4-512
   t^2+456t^{-2}-785\right)+...
\label{vD:C0}
\ee
Corresponding ground state is then given according to \eqref{Eq:psi0} by,
\be
\widetilde \psi_0(x) = 1+\sqrt{pq}\left[16 t^{-1}+8t \left(x+\frac{1}{x}\right)\right]
+pq\left[108 t^{-2}-69+128 \left(x+\frac{1}{x}\right)-\right.
\nn\\\left.
\left(x^2+\frac{1}{x^2}\right)+36t^2\left(1+x^2+\frac{1}{x^2}\right) \right]
+\sqrt{pq}(p+q)\left[16 t^{-1}+8t \left(x+\frac{1}{x}\right) \right]+\cdots\,.
\label{vD:ground:state}
\ee
Now acting with the van Diejen Hamiltonian \eqref{vanDiejen:hamiltonian} on the function above we can check that it is indeed an 
eigenfunction of the operator with the corresponding ground state eigenvalue given by
\be
E_0=1-p-q+...
\label{vD:gs:energy}
\ee
Finally we can also use an expression for the trinion theory derived in \cite{Razamat:2020bix} in order to find $\hat C_0$ defined 
in Section \ref{sec:universality}. The trinion index in the case of equal charges of all moment maps is given by:
\be
{\cal I}(x_1,x_2,x_3)=\frac{(q;q)^2(p;p)^2}{6}\oint\frac{dz_{1,2}}{2\pi i z_{1,2}}\prod\limits_{j\neq i}^3\frac{1}{\GF{\frac{z_i}{z_j}}}
\prod\limits_{i=1}^3\GF{\pq{1}{6}t^{\frac{4}{3}}x_1^{\pm 1}z_i}\times
\nn\\
\GF{\pq{1}{6}t^{-\frac{1}{6}}x_2^{\pm 1}z_i}\GF{\pq{1}{6}t^{-\frac{1}{6}}x_3^{\pm 1}z_i}\GF{\pq{1}{3}t^{-\frac{1}{3}}z_i^{-1}}^6
\,.
\label{vD:trinion}
\ee
Gluing two such trinions along two pairs of $SU(2)$ punctures we obtain a genus-one tube theory. Using this basic building block we can construct higher genus tori leading to 
the following expression for $\hat C_0$:
\be 
\hat C_0=1+ p\, q \left(8 t^3+\frac{8}{t^3}+28 t^2+\frac{28}{t^2}+56 t+\frac{56}{t}+67\right)+...\,.
\label{vD:hatC0}
\ee Note that the coefficient of $q\,p$ is what is expected from decomposition of the  adjoint representation of $\text{E}_8$ into irreps $\SU(8)\times \U(1)$
plus $3$, namely it is ${\bf 248}_{\text{E}_8}+3$.

\

\section{Discussion} 

We have illustrated how using physical input from across dimensions dualities one can generate eigenfunctions and eigenvalues for a variety of elliptic relativistic integrable models.
In principle for any six dimensional SCFT one can associate an integrable model (most straightforwardly if the SCFT when compactified on a circle admits an effective Lagrangian description).\footnote{For more general constructions see {\it e.g.} \cite{Chen:2020jla,Chen:2021ivd,Chen:2021rek,Chen:2023aet,Koroteev:2019gqi,Bullimore:2014awa,Koroteev:2018isw,Gorsky:2021wio,Hatsuda:2018lnv}.}
Thus the method is applicable for this rather large class of models. 
While we have considered several examples in this paper there are still plenty of 
models to be considered. First candidates for this program are models obtained in the compcatifications of the minimal $(D,D)$ conformal matter
and defined on $A_N$ and $C_2$ root systems \cite{Nazzal:2018brc,Nazzal:2023bzu} as well 
as $BC_n$ van Diejen model which is expected to be related to the compactification of the rank $n$ E-string theory \cite{Pasquetti:2019hxf,Hwang:2021xyw}.  
Another important direction of the future research is going beyond the ground states. It would require on one hand more model specific methods ({\it e.g} to deal with ``degeneracies'')
and on the other hand also to develop more sophisticated computational approaches. It would be also interesting to understand whether the functions
derived here admit an independent all order definition. 

\

\noindent{\bf Acknowledgments}:~
We are grateful to Luca Cassia, Alba Grassi and Edwin Langmann for helpful correspondence. 
The research of BN and SSR is supported in part by Israel Science Foundation under grant no. 2159/22, by a Grant No. I-1515-303./2019 from the GIF, the German-Israeli Foundation for Scientific Research and Development,  by BSF grant no. 2018204. The research of AN is supported by he Swiss National Science Foundation, Grant No. 185723.

\bibliographystyle{ytphys}
\bibliography{refs}

\end{document}